\documentclass[10pt,draft]{dis03}
\usepackage{epsf,amsmath}

\begin{document}

\title{UNDERLYING EVENT STUDIES AT CDF}

\author{STEFANO LAMI \thanks{for the CDF Collaboration} \\
The Rockefeller University\\
New York, NY 10021, USA\\
E-mail: lami@fnal.gov }

\maketitle

\begin{abstract}
\noindent We present recent studies about the `underlying event'
which originates mostly from soft spectator interactions.
First Run II data results are compared to published Run I results
and to QCD Monte Carlo models.
\end{abstract}

\section{Introduction}
\begin{figure}[!thb]
\vglue -.1in
\centerline{
\epsfxsize=2.9in\epsfbox{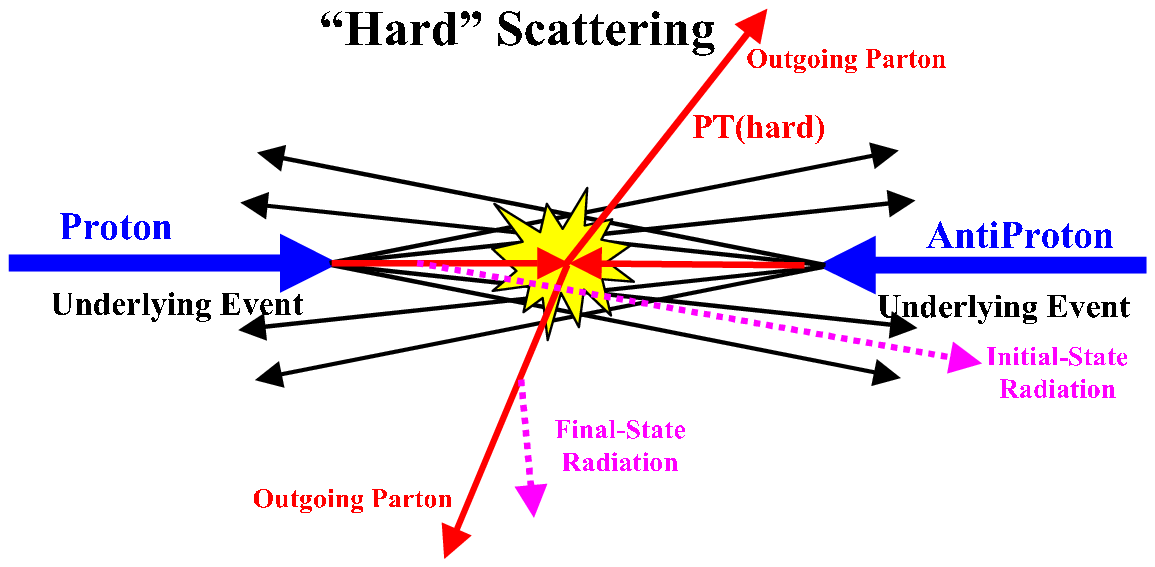}~
\epsfxsize=2.2in\epsfbox{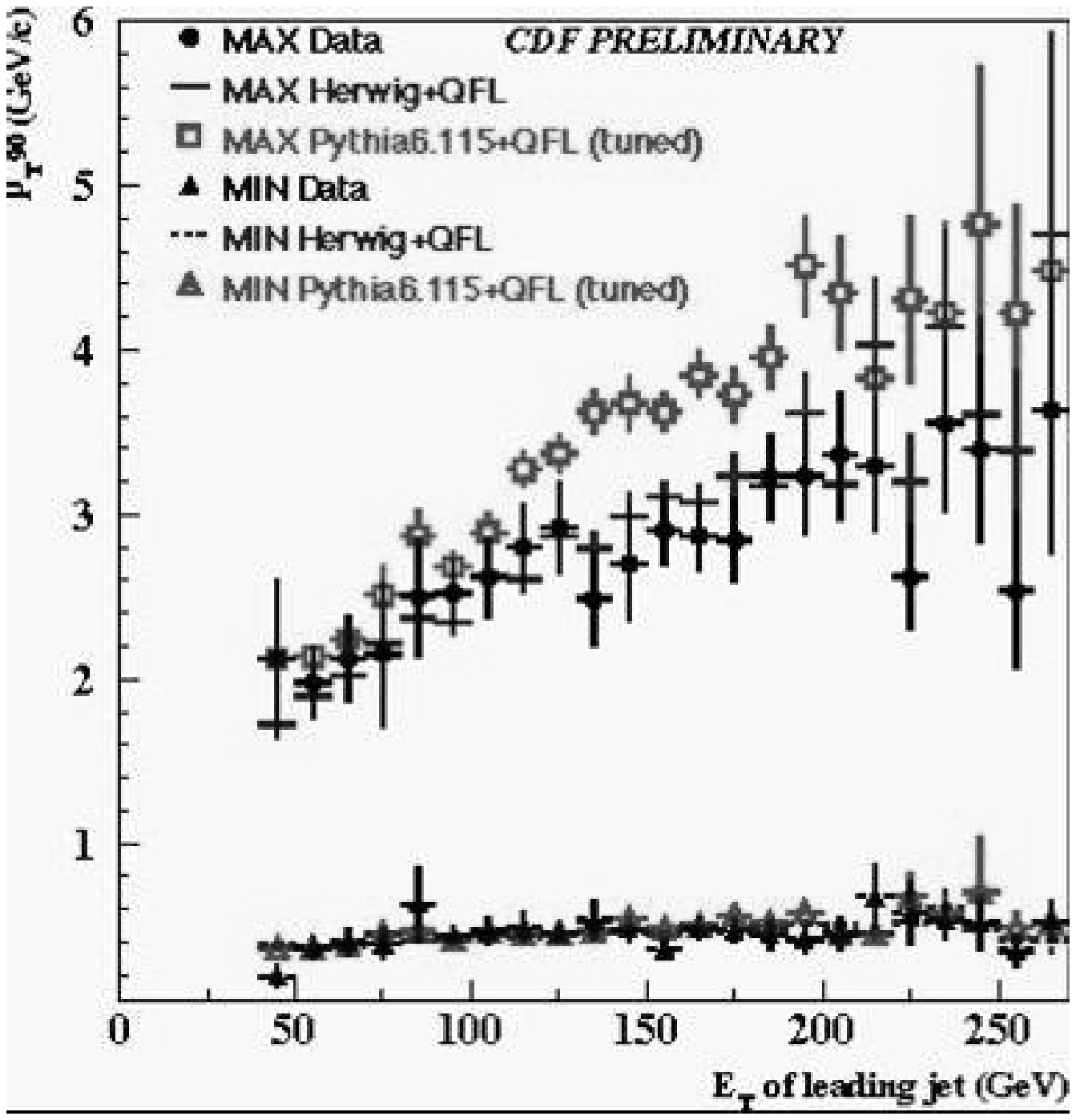}}
\vglue -.1in
\caption[*]{{\it Left:} Illustration of the way QCD Monte Carlo models simulate a proton-antiproton
collision. The `hard scattering' component consists of the outgoing two jets plus initial and final-state radiation.
The `underlying event' is everything except the two outgoing hard scattered jets and consists
of the `beam-beam remnants' plus initial and final-state radiation.
{\it Right:} $\Sigma P_T$ inside the {\it max} and {\it min} cone in jet events as a function of the leading
jet $E_T$.}
\vglue -.1in
\label{hard}
\end{figure}

Most of the inelastic proton-antiproton collisions at the Tevatron 
result in a soft collision with outgoing particles roughly in the same
direction of the initial proton and antiproton, the so-called
minimum bias event.
Occasionally a hard 2-to-2 parton scattering happens, resulting in large
transverse momentum jets whose direction remembers the 2-to-2 hard scattering
subprocess, the two outgoing partons (fig.~\ref{hard} left).
The underlying event (UE) is everything but the two outgoing hard scattered jets, and consists
of the beam remnants plus possible contributions from the initial and final state
radiation. In addition it is possible that a multiple parton scattering, a second semihard 2-to-2
parton-parton scattering, contributes to the UE.

The UE is an unavoidable background whose contribution has to be removed from the jets when we
compare the results with next-to-leading order (NLO) QCD calculations.
The uncertainty on the UE contribution to jets in CDF data is actually the
dominant source of systematic uncertainty for the CDF inclusive
jet cross-section at low energies. 
While it is clear that precise jet measurements require an accurate
modeling of the UE, the understanding of the physics of the UE is quite complicated
and involves both non-perturbative as well as perturbative QCD.
A final solution is not going to happen any time soon and
presently none of the QCD Monte Carlo (MC) models correctly describes the properties
of the UE. Therefore the first question we are trying to answer is whether it would be possible
in the meanwhile
to tune the MC models to better fit the UE.
Another important question is whether minimum bias events could be
a good approximation to the UE. In principle they are different as the UE
has initial state radiation and the beam-beam remnants are color connected
to the hard component of the collision, however we want to estimate
this difference.

The study of the UE involves very low energy particles.
The central tracking system of the CDF detector, immersed into a 1.4 T magnetic field,
measures the trajectory and transverse momentum $P_T$ of charged particles with a good resolution
and a track finding efficiency better than 92\% (for $P_T>$0.5 GeV).
The energy of neutral particles is measured by the calorimeters, but at the low momenta
relevant for this study, the efficiency and resolution of the calorimeter are poor.
In the following, results are presented by using charged particle tracks in the
central region of the CDF detector, and compared to QCD MC predictions.
A particular attention is dedicated to the PYTHIA MC \cite{PYTHIA}
whose UE model includes also multiple parton interactions, which are more likely
to happen in a hard collision.

\section{Run I results: Jet cone analysis}
A first CDF UE analysis \cite{Valeria} studied jet events with energies reconstructed
into a cone of radius $R=$0.7 in the
$\eta$-$\phi$ space, between 50 and 300 GeV.
The transverse momentum $P_T$ of charged particles only is selected, inside
two cones at the same rapidity in the central region (pseudorapidity $|\eta|<$1)
and at $\pm$ 90$^o$ in azimuth from the most energetic jet in the event.
Given the non uniform response of the CDF detector as a function of rapidity,
the requirement of same rapidity is essential.
The two cones are used to study the UE energy in a region transverse to the leading jets.
For each event we label the cone which has the maximum $\Sigma P_T$ as `{\it max cone}' and the
cone with minimum $\Sigma P_T$ as `{\it min cone}'. The difference between the {\it max} and
{\it min} cone provides information on NLO contributions while
the {\it min} cone gives an indication of the level of the UE.
The $\Sigma P_T$ inside the {\it max} and
{\it min} cone is plotted as a function of the leading jet transverse energy $E_T$ in fig.~\ref{hard}(right).
The {\it min} cone has a flat dependence, as independent from the hard interaction,
at a level similar to that found in minimum bias events.
The {\it max} cone increases with the leading jet $E_T$, due to the contribution of a third jet associated
with the hard scattering.
From fig.~\ref{hard}(right) HERWIG \cite{HERWIG} agrees well with these jet data, but does not
reproduce well minimum bias events (not shown), in particular the  $P_T$ distribution
of the tracks, due to the lack of semihard physics. In the  {\it max} cone, PYTHIA is 
somewhat higher than the data for high  $E_T$.

 
\section{Run I results: Charged particle jets}
The second study~\cite{rick_prd} also considers the charged particle components
of jets. It is based on minimum bias  and jet events with a trigger threshold at 20 GeV,
and it compares the data to HERWIG, ISAJET~\cite{ISAJET} and PYTHIA.
Jets are reconstructed with a simple, non-standard algorithm to define clusters of charged particles
with $P_T>$ 0.5 GeV and $|\eta|<$1, within a cone of radius $R=$0.7 in the
$\eta$-$\phi$ space, starting from the most energetic track.
The $P_T$ of this  {\it charged particle jet}  is the scalar sum of the $P_T$'s of all the
charged particles within the jet. With this algorithm even one charged particle can be a jet.

\begin{figure}[!thb]
\vglue -.05in
\centerline{
\epsfxsize=1.5in\epsfbox{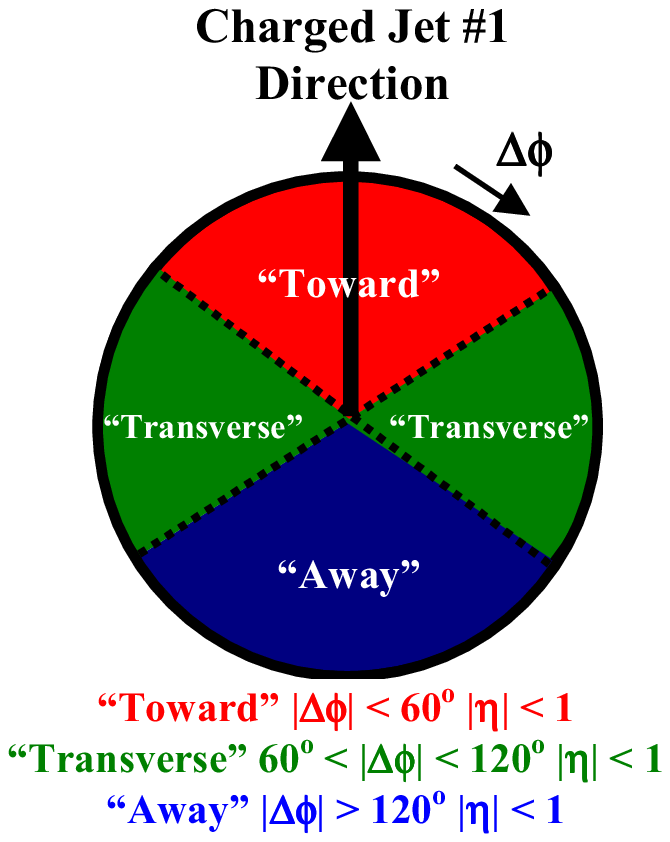}
\epsfxsize=3.2in\epsfbox{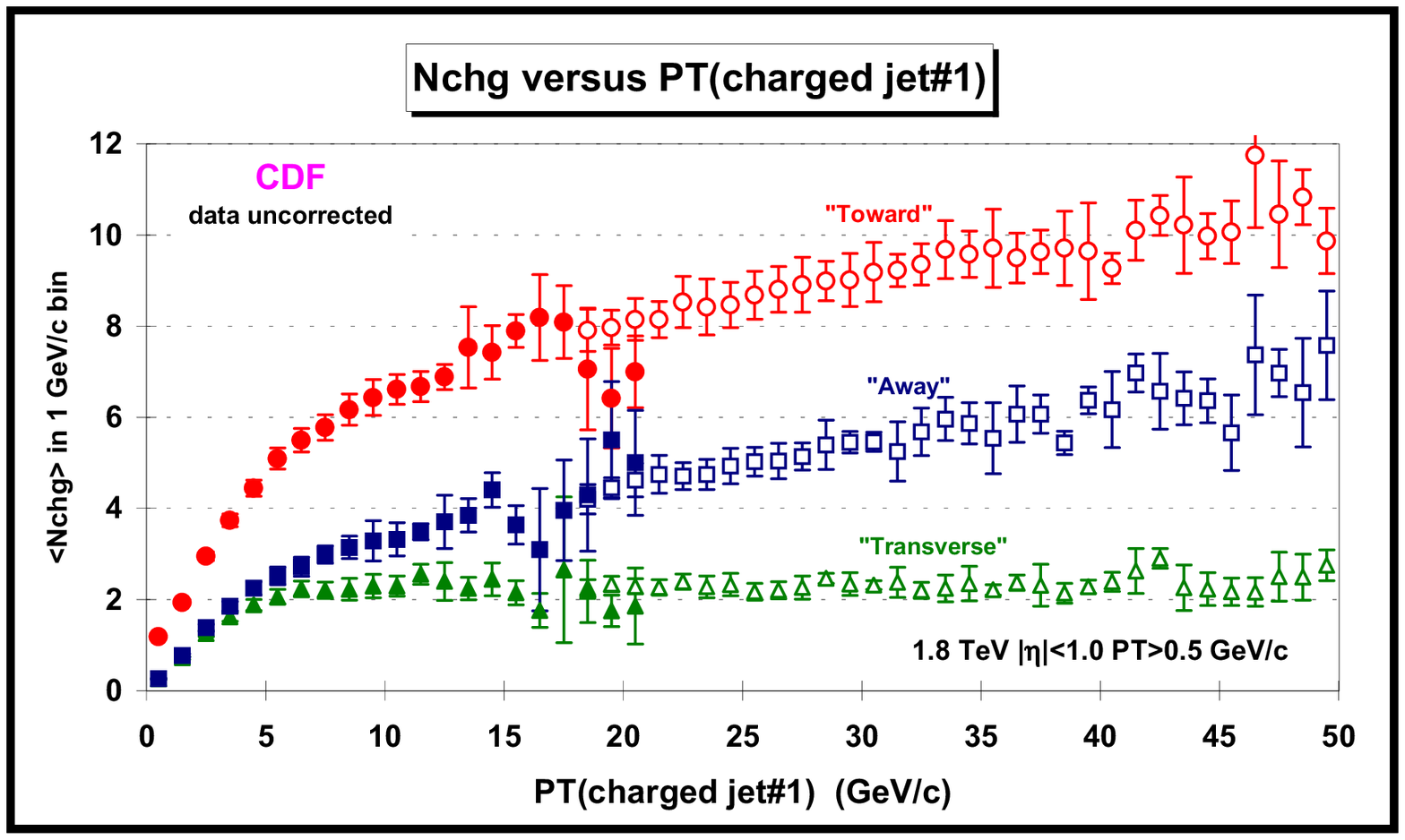}
}
\vglue -.11in
\caption[*]{{\it Left:} Three regions in azimuthal angle relative to the leading jet direction are defined to
separate the different hadronic activities in the event. {\it Right:} Average number of charged particles 
in the three different regions as a function of the leading {\it charged particle jet} $P_T$. Solid points are
for minimum bias events while open points are for jet events.}
\vglue -.05in
\label{ch_jet}
\end{figure}

Given the direction of the leading  {\it charged particle jet}, we define three regions in $\phi$
to subdivide the different hadronic activities in the event (fig.~\ref{ch_jet}).
Based on the topology of hard scattering jet  events, we define the {\it toward region} as the one
which contains the leading {\it charged particle jet}, and covers the region $\Delta\phi=\pm 60^o$
from the leading jet direction.
The {\it away region} on the average contains the away side jet, as the two large transverse momentum
jets are roughly back to back in azimuthal angle.
The transverse region is perpendicular to the plane of the hard 2-to-2 scattering and is very sensitive to the
UE component of the QCD MC models.
All three regions have same size, 4$\pi$/3, in $\eta$-$\phi$ space.

Fig.~\ref{ch_jet}(right) shows the average number of charged particles in the three different regions
as a function of the leading {\it charged particle jet} $P_T$, for minimum bias (solid points) and jet (open points)
events.
In this and following plots the data are not corrected up for the track finding efficiency. Rather, MC
events are corrected down. For the selected $P_T$ and $\eta$ region, these corrections are small and essentially independent of $P_T$ and  $\eta$, which is why the study uses only charged particles in this limited region.

\begin{figure}[!thb]
\vglue -.15in
\centerline{
\epsfxsize=2.6in\epsfbox{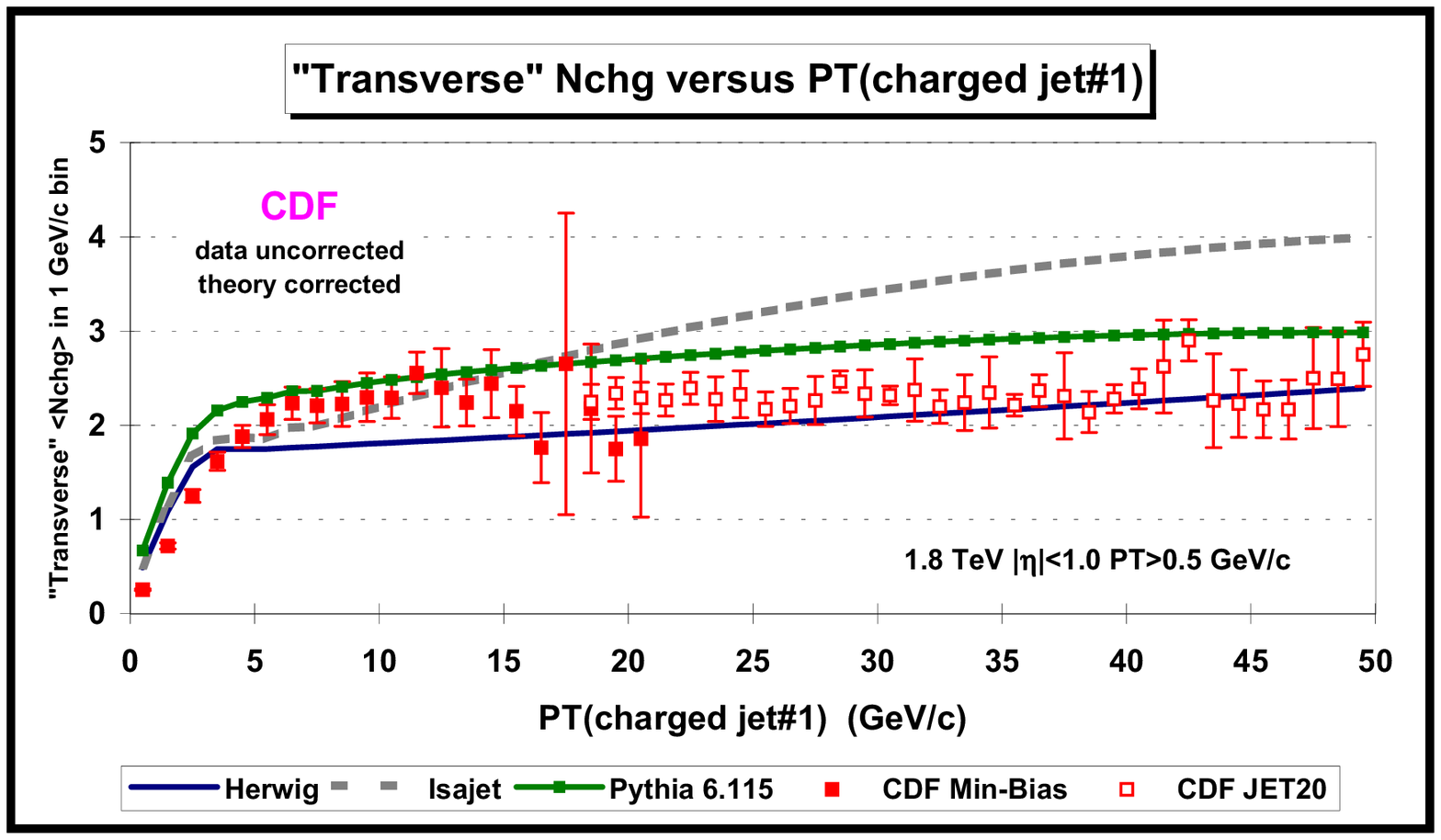}
\epsfxsize=2.6in\epsfbox{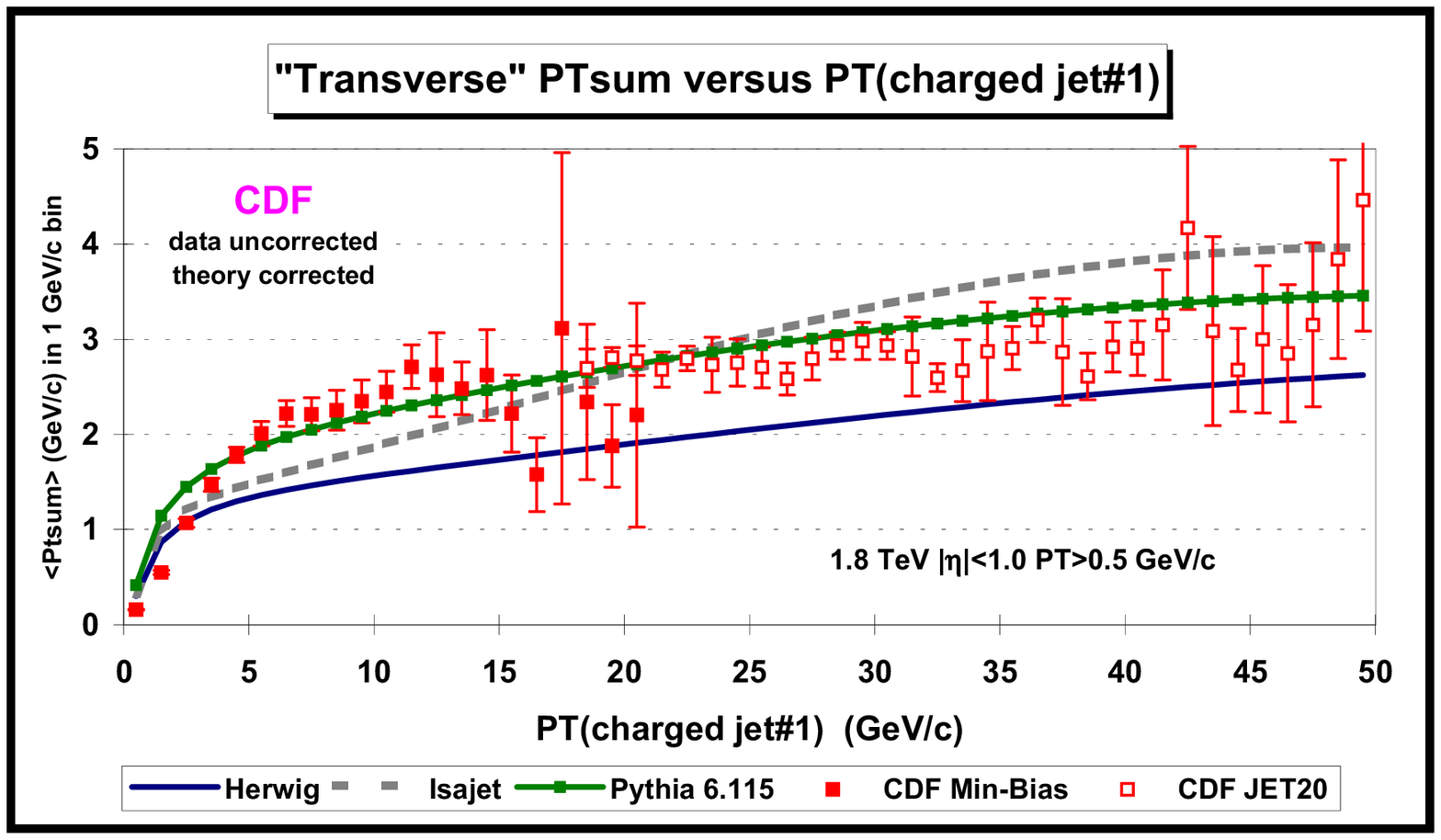}
}
\vglue -.1in
\caption[*]{Average number of charged particles ({\it Left}) and their total $P_T$  ({\it Right}) in the transverse  region as a function of the leading {\it charged particle jet} $P_T$. Solid points are
for minimum bias events while open points are for jet events.}
\vglue -.05in
\label{prd}
\end{figure}

We focus on the transverse region which is very sensitive to the UE. For central tracks with $P_T>$0.5 GeV,
fig.~\ref{prd}(left) shows better how the jet events in the transverse region are a factor
2 more active than an average minimum bias event, and form a plateau above 5 GeV.
Fig.~\ref{prd}(right) shows  the total $P_T$ of charged particles in the transverse  region as a function of the leading {\it charged particle jet} $P_T$.
If we split the transverse regions in two regions, {\it max} and {\it min} according to their total charged particle
$P_T$ on an event by event basis, 
results are in very good agreement
with the max/min cone study of previous section, once the data are normalized to the different area.

\section{QCD Monte Carlo models}

Fig.~\ref{prd} also shows a comparison of CDF Run I data to some QCD MC models.
In general, the agreement between the predictions for the transverse region and data is poor.
HERWIG does not predict enough activity in the transverse region. ISAJET, on the contrary, 
has a lot of activity but with the wrong dependence on the the leading jet $P_T$.
PYTHIA with default parameters gives a poor description of the UE. The sensitivity to the choice
of different parton distribution functions has been studied and showed a small effect.

From fig.~\ref{prd} it seems then clear that, even if the approximation of the UE by using
minimum bias events has to be taken very carefully due to the doubled activity of the UE in jet events,
nevertheless minimum bias events do a better job than current predictions when used to
represent the evolution of the charged particle jets with the event energy. They provide
a smooth continuation at very low energies of the high $P_T$ charged jets.

However, we found out that PYTHIA can be tuned to fit pretty well CDF Run I data by enhancing the multiple
parton interactions, 
tuning the relative parameters for the impact parameter and a double
gaussian matter distribution~\cite{rick}.
This tuned version of PYTHIA can also describe well the transition region between soft and hard collisions,
as shown in fig.~\ref{run2} which will be discussed in the next section.
Different tunings of PYTHIA regarding the initial state radiation show that the increased activity in the UE in a hard
scattering over a soft collision cannot be explained by initial state radiation.
Whether the multiple parton interaction approach is correct or we simply need to improve how MC models handle the 
beam-beam remnants, the multiple particle interactions give a natural way to explain the increase of
activity in the UE of a hard scattering, where its probability is higher.

\section{The `Underlying Event' in Run II}
The Tevatron Collider is currently running at a larger center of mass energy, but the UE is expected
to change negligibly between $\sqrt{s}=$1.8 TeV of Run I and the present Run II $\sqrt{s}=$1.96 TeV.
We present here preliminary results for the {\it charged particle jet} study. 
CDF has a new central tracking system with similar performance to Run I for the resolution of charged
particle $P_T$. The quality cuts for the track selection are the same, and like in Run I
we do not consider events with two or more vertices.

\begin{figure}[!thb]
\vglue .02in
\centerline{
\epsfxsize=2.6in\epsfbox{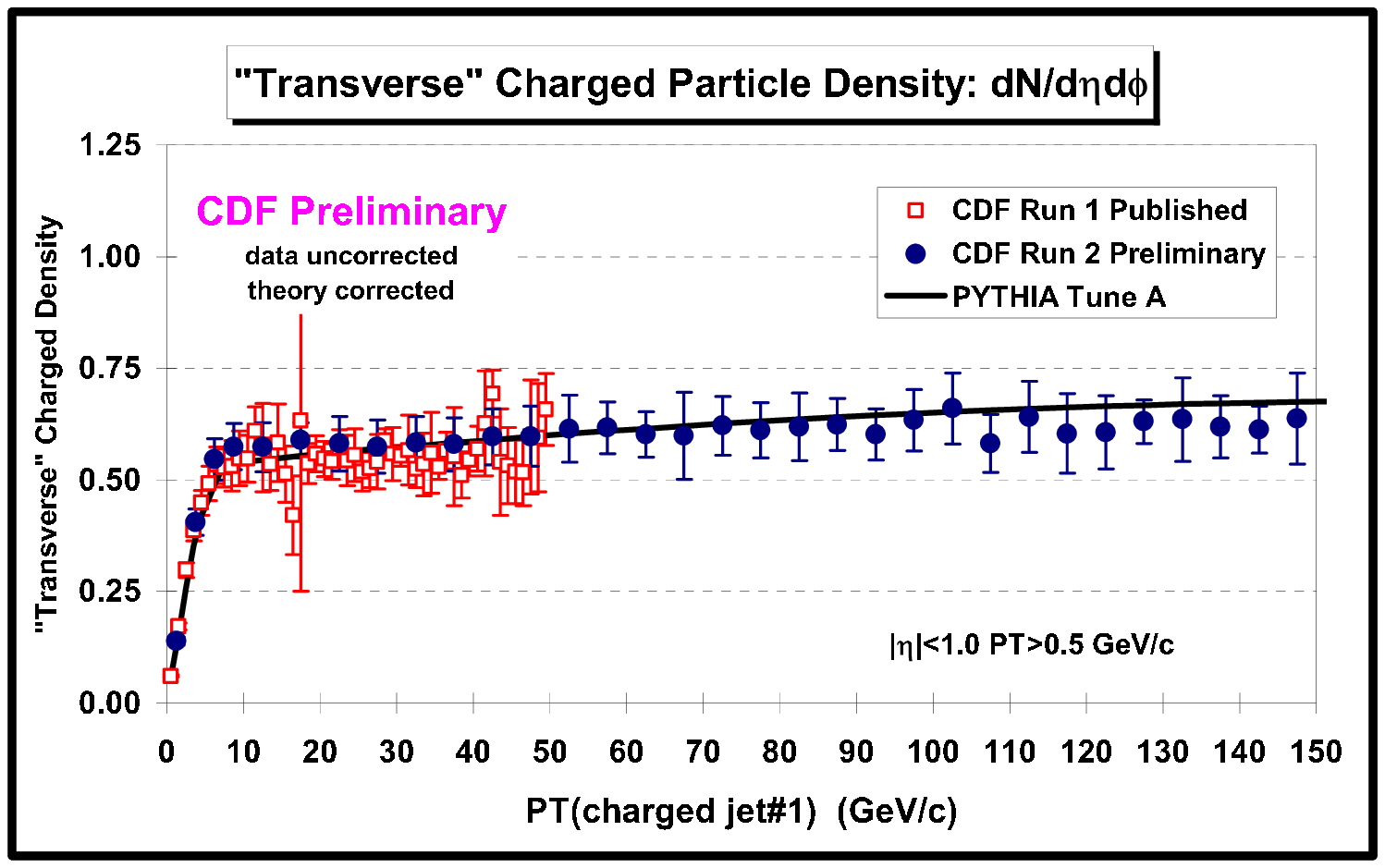}
\epsfxsize=2.6in\epsfbox{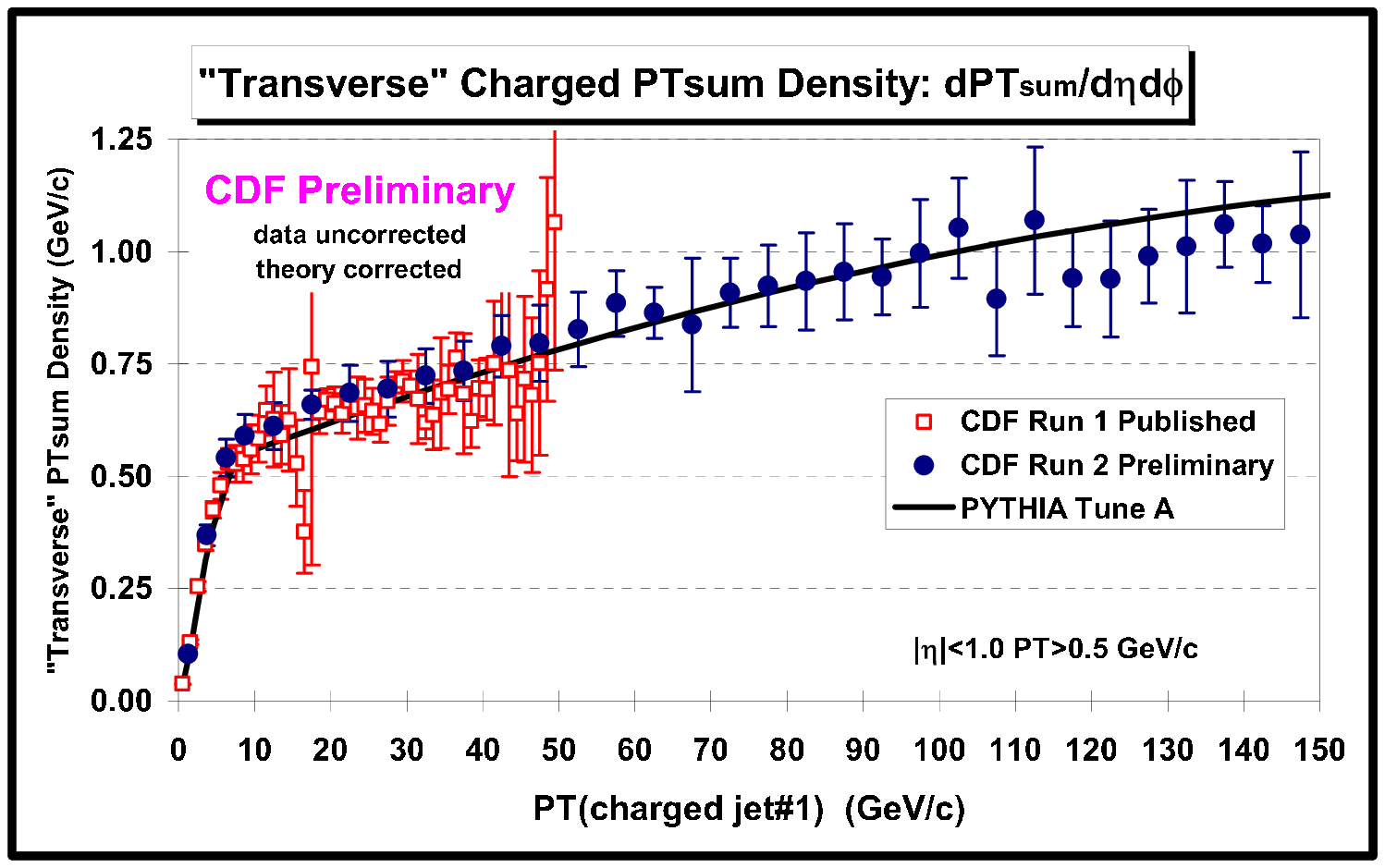}
}
\vglue -.1in
\caption[*]{Run II data (solid points) on the average density of charged particles ({\it Left}) and on the average
charged $\Sigma P_T$ density  ({\it Right}) ($P_T>$0.5 GeV, $|\eta|<$1) in the transverse  region as a 
function of the leading {\it charged particle jet} $P_T$. They are compared with the published
CDF Run I data (open points), while the theory curve corresponds to PYTHIA Tune A.}
\vglue -.1in
\label{run2}
\end{figure}

Minimum bias and jet data are combined together in fig.~\ref{run2} and compared
to Run I results as well as to the PYTHIA version (6.206) tuned on Run I data.
The left plot shows the average density of charged particles in the transverse region,
i.e. per unit of pseudorapidity and azimuthal angle, $dN/d\eta~d\phi$. 
The right plot shows the average charged particle total $P_T$ density,
$dP_T^{sum}/d\eta~d\phi$, as a function of the  leading {\it charged particle jet} $P_T$. 
There is an excellent agreement between Run I and Run II data and the tuned version of PYTHIA.

\section{Conclusions}
\vspace*{-.1in}
Combining minimum bias and jet CDF data provides a good sample to study the evolution of the
UE from very soft collisions to very hard scatterings.

MC models with default parameters are inadequate in describing the UE in CDF and this has led to improved tuning.
Tuned PYTHIA (with multiple parton interactions) does a good job in describing the UE in CDF data.

Run I and Run II data show an excellent agreement for charged particles. The UE is the same in both samples,
but now we can study the evolution out to higher energies.

\section*{Acknowledgements}
\vspace{-.1in} 
Many thanks to Rick Field for his precious advice, and to the workshop organizers for the warm
hospitality.

\end{document}